\pdfoutput=1
\documentclass[aps,pre,twocolumn,groupedaddress]{revtex4-1}
\usepackage{graphicx}
\usepackage{lipsum}  % For dummy text
\usepackage[utf8]{inputenc}
\usepackage{mathtools}

\usepackage{float}
\usepackage[margin=1in]{geometry} % Adjust margins as needed
\usepackage[usenames,dvipsnames]{xcolor}
\usepackage{parskip} % Comment this out or adjust manually

\usepackage{relsize}
\usepackage{amsmath}    % need for subequations
\usepackage{amssymb}
\usepackage{bm}

\usepackage{hyperref}
\usepackage{latexsym}
\usepackage{verbatim}
\usepackage{color}
\usepackage{xcolor}
\enlargethispage{\baselineskip}

\def\beq{\begin{equation}}
\def\eeq{\end{equation}}

\def\beq{\begin{equation}}                          
\def\eeq{\end{equation}}                          
\def\bea{\begin{eqnarray}}                          
\def\eea{\end{eqnarray}}

\DeclareRobustCommand{\uvec}[1]{{%
  \ifcsname uvec#1\endcsname
     \csname uvec#1\endcsname
   \else
    \bm{\hat{\mathbf{#1}}}%
   \fi
}}

\preprint{}

\begin{document}

%%%%%%%%%%%%%%%%%%%%%%%%%%%%%%%%%%%%%%%%%%%%%%%%%%%
%                                 TITLE & ABSTRACT
%%%%%%%%%%%%%%%%%%%%%%%%%%%%%%%%%%%%%%%%%%%%%%%%%%%

\title{From Flocking to Condensation: Collective Dynamics in Binary Chiral Active Matter}
\author{Divya Kushwaha}
\email{divyakushwaha.rs.phy22@itbhu.ac.in}
\affiliation{Indian Institute of Technology (BHU) Varanasi, India 221005}

\author{Shradha Mishra}
\email[]{smishra.phy@itbhu.ac.in}
\affiliation{Indian Institute of Technology (BHU) Varanasi, India 221005}
\date{\today}
\begin{abstract}
{Many microswimmers are inherently chiral, and this chirality can introduce fascinating behaviors in a collection of microswimmers. The dynamics become even more intriguing when two types of microswimmers with distinct chirality are mixed. Our study examines a mixture of self-propelled particles with opposite chirality, investigating how the system's characteristics evolve as the magnitude of chirality is varied. In weakly chiral systems, the particles exhibit similar behavior, leading to a globally flocking phase where both types of particles are well-mixed. However, in an intermediate range of chirality, the condensates of different particles are formed as a result of a competition between chirality and self-propulsion. This competition results in interesting phases within the system. We explore the characteristics of these distinct phases in detail, focusing on the roles of self-propulsion speed and chirality.}
\end{abstract}
\maketitle
\section{Introduction}
Active matter is a fascinating area of study \cite{sumpter2010collective,vicsek2012collective,mishra2024directional,lavi2024dynamical} that focuses on self-propelled entities such as active colloids \cite{bechinger2016active}, insects swarming in large groups \cite{cavagna2017dynamic}, fish displaying schooling in the ocean \cite{ward2008quorum}, and birds flocking in the sky \cite{ballerini2008interaction}, groups of animals at the macroscopic scale \cite{cavagna2014bird}, {\em etc.} These entities move due to internal mechanisms such as chemical reactions, cilia, flagella, and internal motors that utilize the input from the surrounding mediums. Understanding collective behavior in active matter is crucial to interpret collective phenomena. Flocking, which refers to the collective motion of organisms, is also observed in inanimate macroscopic systems, such as active granular rods \cite{kumar2014flocking}. This widespread phenomenon often involves transitions from disordered to long-range ordered states, influenced by factors such as density and noise \cite{r1}. \\
The dynamics of active agents can be categorized into two main types: one involves nearly straight movement with random directional changes, creating a sense of self-propulsion without any perception of rotation, as seen in active colloids like Janus particles, molecular motors \cite{julicher1997modeling}, natural low-Reynolds-number swimmers \cite{najafi2004simple,ebrahimian2015low,bennett2013emergent}, active particles on periodic lattice \cite{chamolly2017active} and polar active granular particles \cite{zhang2017janus}. 
Particles with a sense of rotation or chiral particles characterize the second type. Systematically, particles can tend to rotate clockwise $(CW)$ or anticlockwise $(ACW)$. The motion is circular in two dimensions and helical in three dimensions \cite{lowen2016chirality,friedrich2016hydrodynamic}. These chiral particles possess a characteristic self-propulsion velocity and a chirality that determines the rate of directional change in the absence of fluctuations. The radius of their trajectories is determined by the ratio of self-propulsion velocity to chirality \cite{caprini2024self}. This type of chiral motion is observed in various systems, including biological circle swimmers such as bacteria \cite{diluzio2005escherichia}, E.coli \cite{berg1990chemotaxis}, sperm cells \cite{riedel2005self}, artificial microswimmers of L shape \cite{kummel2013circular}, and magnetotactic bacteria in rotating external fields \cite{erglis2007dynamics}. 

Although most of the research on bacterial swarming and collective swimming has focused on homogeneous systems \cite{harshey2003bacterial,harshey2003bacterial,sokolov2007concentration,copeland2009bacterial}, which involve agents having identical features. The natural world encounters mixtures of particles of different chiralities \cite{schumacher2017semblance,jolles2020role, ao2015diffusion}. These differences arise from variations in their size \cite{peled2021heterogeneous}, diverse orientations, and behavior of mixed species \cite{benisty2015antibiotic,zuo2020dynamic}. Complex systems are naturally diverse, as individuals vary in their characteristics, environmental reactions, and interactions. The collective behavior of such diverse systems has been thoroughly examined, revealing complexities that differ from those found in uniform systems.\\
Recent experiments have explored mixtures of two types of chiral particles, revealing fascinating collective behaviors. In some studies, chiral particles with clockwise $(CW)$ and anticlockwise $(ACW)$ rotations are generated using confined geometries with polarized walls \cite{barois2020sorting}. In the case of run-and-tumble particles, a high tumbling rate can induce chiral motion, causing bacteria to behave like chiral swimmers \cite{ariel2018collective}. This phenomenon also demonstrates a method for sorting particles based on their physical properties using asymmetric barriers without external driving forces. \cite{reichhardt2013dynamics}. In one of our previous works \cite{eswaran2024synchronized}, we also found the synchronized rotation of active Brownian particles due to chemotaxis, which is another way of introducing the chirality in the dynamics of particles due to the medium.
Reversible pattern formation and edge transport have also been studied in mixtures of active chiral and passive disks, where the degree of segregation depends on orbit radius and density \cite{reichhardt2019reversibility}. Additionally, probe particles driven through chiral fluids exhibit transverse Hall-like motion, revealing the role of chirality in microrheological response and jamming behavior \cite{reichhardt2019active}. 
Some significant studies have delved into macroscopic colony growth and spatial mixing between populations \cite{deforet2019evolution,levis2019activity,liebchen2017collective,levis2019simultaneous,ventejou2021susceptibility, mecke2024emergent}. \\
In this study, we consider a minimal model of point-like chiral active particles interacting through local velocity alignment in a binary mixture with equal and opposite chirality. Our model is inspired by the Vicsek model\cite{vicsek1995novel}, which introduced alignment-based interactions among self-propelled particles as a paradigm for collective motion. However, unlike the Vicsek model—which considers homogeneous particles aligning their velocity—we introduce heterogeneity through a binary mixture of oppositely chiral particles. We introduce heterogeneity through a binary mixture of oppositely chiral particles. We introduce two types of chiral particles, each constituting half of the total population, with one group exhibiting clockwise $(CW)$ rotational tendency and the other anticlockwise $(ACW)$ tendency. This inherent rotation creates heterogeneity within the system. Our primary aim is to understand how these particles interact when they encounter those with particles of opposite nature, particularly under varying parameters {\em like}: self-propulsion speed ($v_0$) and chirality ($\Omega$). At low $\Omega$, the system shows mixed flocking behavior where particles exhibit only slight rotational tendencies. As $\Omega$ increases, $CW$ and $ACW$ particles separate, forming dense clusters aligned with their rotational directions, resulting in two distinct clusters at moderate $\Omega$ values.
Interestingly, particles remain homogeneously mixed at large chirality at low self-propulsion speeds ($v_0 = 0.001$), rotating uniformly on their axes rather than separating. This phenomenon occurs because, as $\Omega$ increases, particles show biased behavior based on their orientation; positive and negative chiral particles tend to move in $CW$ and $ACW$ trajectories, but the self-propulsion speed is not strong enough to separate them.
The self-sorting of the chiral species in the mixture of chiral species is also observed in the experiment of \cite{gubitz2001chiral}, where the active particles segregate based on chirality without the need for chemical reactions. The clusters of segregated chiral particles of the same type observed in our study resemble the structures seen in experiments \cite{pi2022baffle,sirota2010genome, denk2016active}.
The phase separation kinetics show the algebraic growth with a growth exponent close to $1/4$. The same as for other active systems \cite{pattanayak2021domain, pattanayak2020speed, wittkowski2014scalar}. This study highlights the intricate interplay between self-propulsion speed and rotational bias due to chirality, revealing various complex behaviors in chiral active systems.\\
The rest of the article is structured as follows: Section \ref{secII} introduces the model for the dual chiral active particle system. The results and discussion of our system are detailed in Section \ref{secIII}. Section \ref{subsec:Dynamic} explores the dynamic phases of particles. Finally, Section \ref{secIV} discusses our results and conclusions.
 \section{Model and Numerical Details\label{secII}}
\label{sec:Model}
Our system comprises two chiral particles, each containing an equal number of $N_1=N_2=N/2$ particles. These particles possess the same magnitude of chirality with opposite signs. They are distributed across a two-dimensional, off-lattice plane within an $L \times L$ square box with periodic boundaries. The particles are represented as point particles and interact within a defined radius denoted as R. Initially, these particles are randomly dispersed in terms of both position $\textbf{r}_i$ and direction ($\theta_i$ ranging from $[-\pi,\pi]$), each having a constant self-propelled speed $v_0$. The temporal update of position and direction is done at an interval of $\Delta t$. The particle's position is updated as follows:
\begin{equation}
  {\textbf r_i}(t+\Delta t)= {\textbf r_i}(t) + {\textbf v_i}(t)\Delta t
   \label{eq1}
\end{equation}
In the given equation, ${\textbf r_i}(t)$ denotes the positions of the $i^{th}$ particle at time $t$, where $i$ can run from $i=1,2,.......N$.
\begin{figure}[h]
\centering
\includegraphics[width=4.3cm, height=4cm]{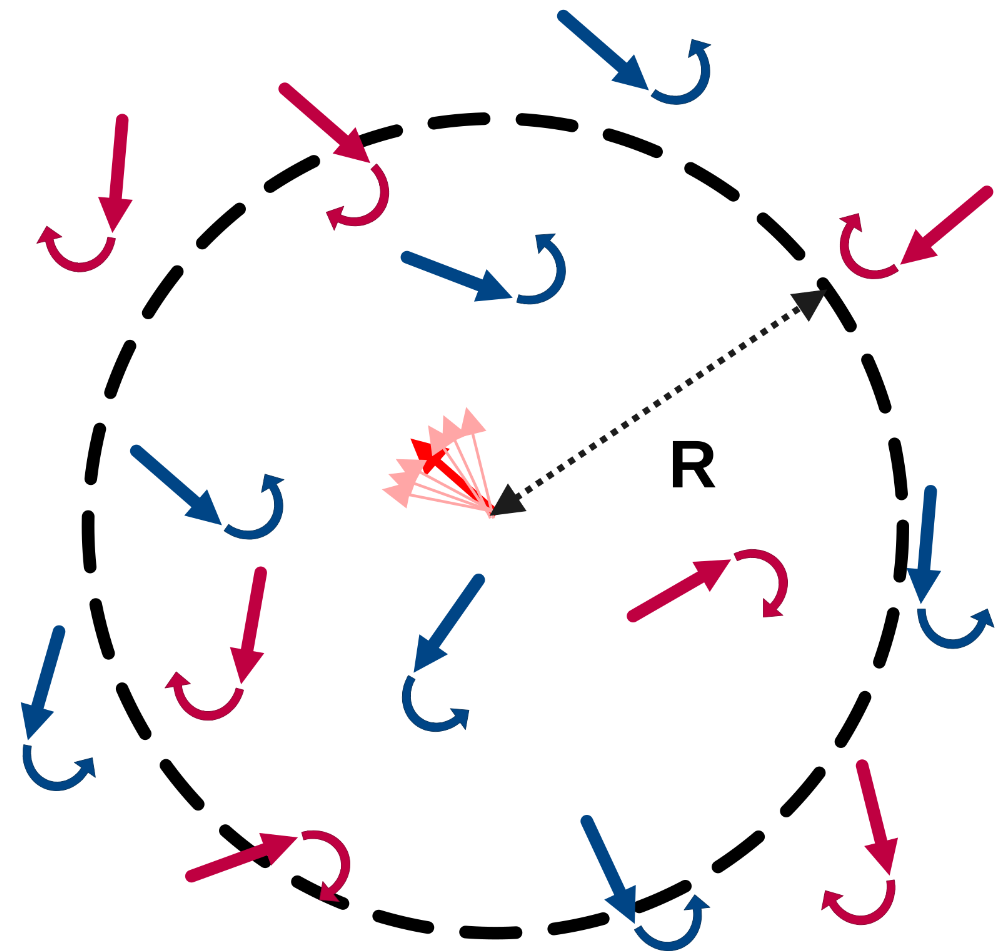}
\caption{(color online) A schematic representation of the model showing two types of chiral particles with their orientation directions: red for $\Omega^+$ (clockwise rotation, $CW$) and blue for $\Omega^-$ (anticlockwise rotation, $ACW$). The red arrow in the center of the dashed circle is the $i^{th}$ particle, and the dotted line represents the radius of the interaction range $R$. The light red arrows show the effect of random noise $\Delta \theta$. The curved arrows with each particle represent the chiral nature of particles.}
\label{fig1}
\end{figure}
${\textbf v_i}(t)\equiv (v_0\cos(\theta_i), v_0 \sin(\theta_i))$, represents the velocity of an $i^{th}$ particle at time $t$. Here, $\theta_i(t)$ is its orientation, and each particle moves with a constant self-propulsion speed $v_0$. The particles interact locally within their interaction radius $R$ as shown in Fig.\ref{fig1}. To incorporate the chiral nature of the particles, the direction of the  $i^{th}$  particle is updated according to:
 \begin{equation}
  \theta_i (t+\Delta t) =  \langle \theta_j (t) \rangle_{i R} + \Omega \Delta t + \Delta \theta 
   \label{eq2}
 \end{equation}
 The $\langle \theta_j (t) \rangle_{iR}$ represents the average of the directions of all the $j$ particles within a defined interaction radius $R$ of the $i^{th}$ particle. This average direction is calculated by considering the orientations of all particles within this interaction radius. Mathematically, it is expressed as:
\begin{equation*}
 \langle \theta_j (t) \rangle_{i R} = tan^{-1}\Bigg[\dfrac{\langle sin\theta_j (t) \rangle_{i R}}{\langle cos\theta_j (t) \rangle_{i R}}\Bigg]
\end{equation*}
In this context, $\langle \sin\theta_j (t) \rangle_{i R}$ and $\langle \cos\theta_j (t) \rangle_{i R}$ denote the average values of $\sin\theta$ and $\cos\theta$, respectively, for all particles within the interaction radius $R$ of the $i^{th}$ particle.
Chirality is introduced using the parameter $\Omega$, where a positive value (denoted as $\Omega^{+}$) signifies $CW$ rotation and a negative value (denoted as $\Omega^{-}$) represents $ACW$ rotation. The magnitude of the chirality is given by $|\Omega^{\pm}|=\Omega$.  $\Delta \theta$ represents the angular noise, which is modeled as $\Delta \theta=\eta \xi_i(t)$, $\eta$ represents the noise amplitude  and  $\xi(t)$ is a random  variable with  zero mean $<\xi_i(t)>  = 0$, delta correlated $<\xi_i(t_1)\xi_j(t_2)> = \delta_{ij} \delta(t_1 -t_2)$ and uniformly distributed in $[-\pi,\pi]$. This directional update ensures that the particles demonstrate chiral behavior while considering the average direction of their neighbors within the interaction radius. The noise strength $\eta$ is set to $0.2$, so the system is in the ordered state for a zero chirality $\Omega = 0.0$. The time difference for updating particles' position and orientation is fixed to $\Delta t=1.0$. The system is studied for $v_0\in(0.001,0.8)$ and $\Omega \in (0.001,0.5)$.
\begin{figure*}[!hbtp]
\centering
\includegraphics[width=13cm, height=20.5cm]{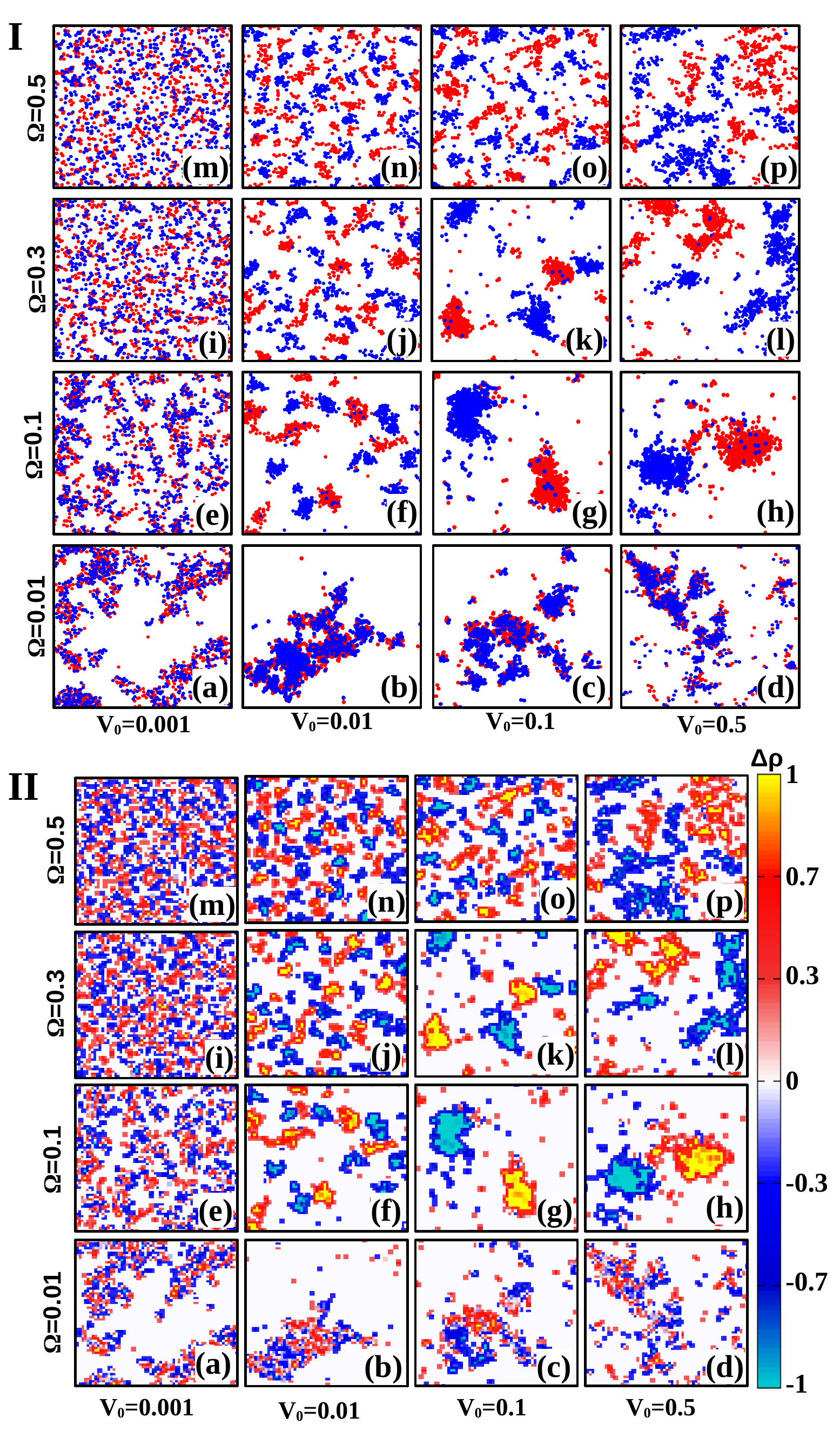}
\caption{(color online) (I) The figure illustrates the snapshots of the system for $L=40$ and at time $t = 2^{17}$. The red dots correspond to particles with positive chirality, denoted by $\Omega^+$, and the blue dots represent particles with negative chirality $\Omega^-$. These particles are distributed across the coordinate plane, visualizing their spatial arrangement. (II) This figure shows the density difference $\Delta \rho$ for the same system parameters as for (I). The color map indicates the $\Delta \rho$ range: blue regions where $\rho^-$ exceeds, red regions where $\rho^+$ dominates, and white regions are empty. Both panels are depicted in the parameter space of chirality $\Omega$, varying from $0.01$ to $0.5$, and self-propulsion velocity $v_0$, ranging from $0.001$ to $0.5$.}
\label{fig2}
\end{figure*}
Additionally, the interaction radius is fixed at $R=1$, and the total particle density is maintained at $\rho = \dfrac{N}{L^2} = 1.0$ $(N = 400-10000, L=20-100)$, with $CW$ and $ACW$ particles each having  density $0.5$. We let the system evolve, starting with the random homogeneous mixed state of both types of particles, and the position and orientation of particles are updated using equations \eqref{eq1}, \eqref{eq2}. Once all the $N$ particles' positions and orientations are updated, one simulation step is counted. The kinetics of the system are studied for a total of $t=2^{16}$ time steps, and steady-state results are obtained by total simulation steps of $2^{17}$, with time average over the last $2^{15}-2^{17}$ time steps. Further, data averaging for the system's kinetics over $200-500$ independent realizations is performed. The steady-state results are obtained by ensemble averaging data over $10$ independent realizations.
\section{Results and Discussion\label{secIII}}
\begin{figure*}[!hbtp]
\centering
\includegraphics[width=15cm, height=12cm]{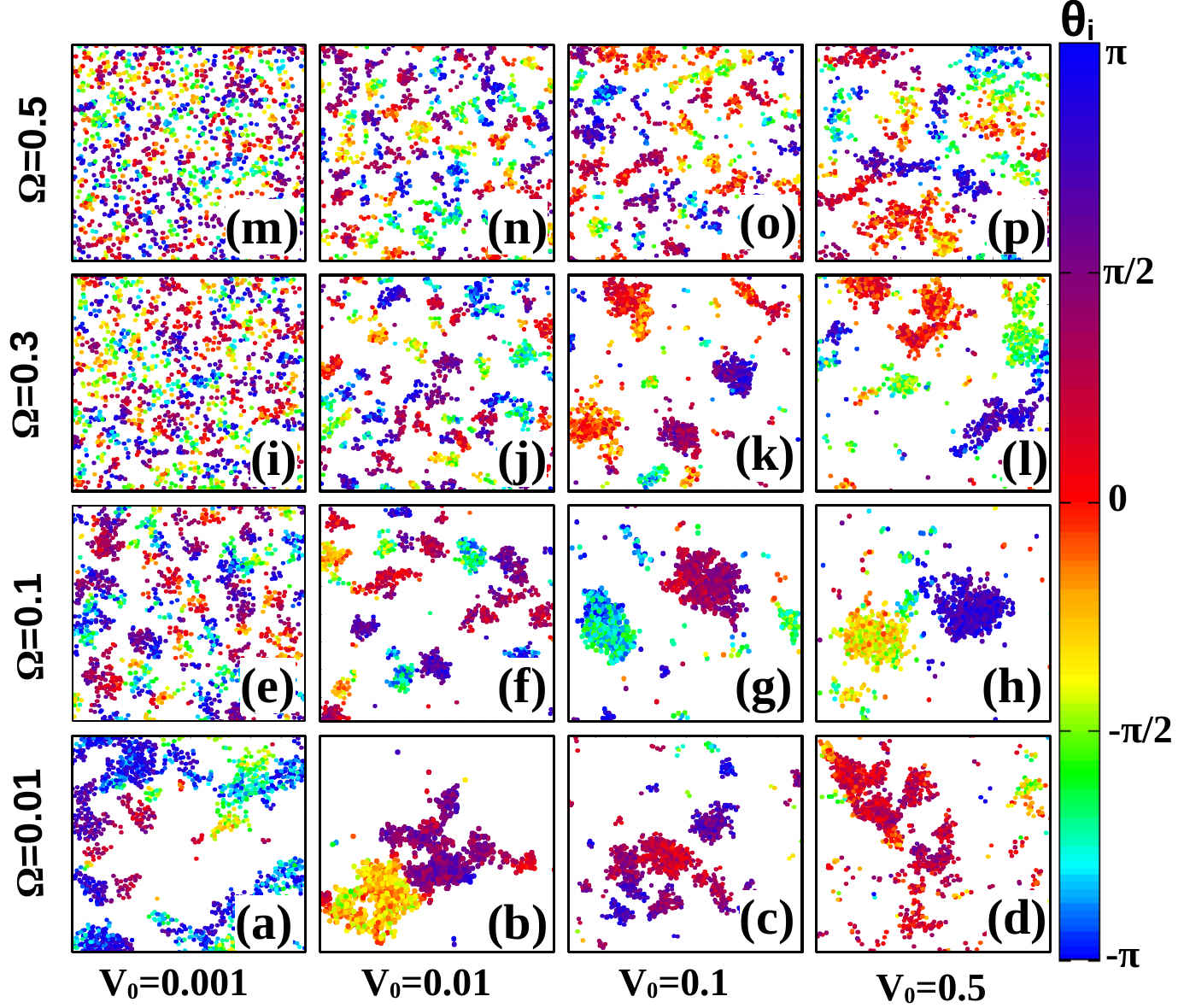}
\caption{(color online) Shows the snapshots for the orientation of particles. 
In this figure, the color represents the direction of particles $\theta_i$ for different $\Omega$ and $v_0$. The parameters are the same as in Fig. \ref{fig2}(I)}
\label{fig3}
\end{figure*}
Our study focuses on two distinct parameters: the self-propulsion speed  $(v_0)$ and the magnitude of chirality $(\Omega)$. By systematically varying these parameters, we can observe the emergence of various phases within the system characterized by fixed noise $(\eta)$ and density $(\rho)$. 
\subsection{Steady state characteristics}
\label{subsec:characteristion}
We first show the snapshots of the system in the steady state at time $t=2^{17}$. In Fig.\ref{fig2}(I), we show the location of two types of particles with two different colors (the red and blue colors represent the $CW$ and $ACW$ particles, respectively), and Fig.\ref{fig2}(II) shows the local density difference between the two types of particles. The local density of both types of particles $\rho^{\pm}({\bf r})$ in the system is obtained by dividing the whole system into coarse-grained square cells of size $0.625 \times 0.625$ and counting the density of particles in each cell centered around point ${\bf r}$ in the system. Further, the normalized density difference of the two types of particles is defined as $\Delta \rho({\bf r}) = \frac{\rho^{+}({\bf r}) - \rho^{-}({\bf r})}{\rho^{+}({\bf r}) + \rho^{-}({\bf r})}$. In Fig.\ref{fig2}(II), we show the snapshots of the system for the density difference $\Delta \rho({\bf r})$ shown in the color bar. The white regions in Fig.\ref{fig2}(II) represent the empty regions. The positive and negative values of $\Delta \rho$ indicate that most particles are of $CW$ and $ACW$ in nature, respectively. In Fig.\ref{fig3} we show the orientation $\theta_i \in (-\pi, \pi)$ of particles at the same time as for Fig.\ref{fig2}. Now we will explain our observations from Figs.\ref{fig2}(I-II) and \ref{fig3}. \\
In the lower chirality regime $\Omega=0.01$, the particles remain mixed irrespective of their speed, as is evident from the density snapshots in Figs.\ref{fig2}II(a-d) and the particle snapshots in Figs.\ref{fig2}(I)(a-d). However, there is a change in global alignment among the particles in each participating cluster as the self-propulsion velocity is tuned. For smaller magnitudes of self-propulsion, the particles rotate about their axis before they go beyond their interaction region, which hinders their chance to interact with new neighbors. The system remains nearly homogeneous and has weak orientational ordering, as can be seen by multiple colors in the orientation plot for small $v_0 = 0.001$ and $0.01$ in Figs.\ref{fig3}(a,b). However, as $v_0$ increases, self-propulsion dominates the rotation due to chirality, enhancing particle interaction and giving rise to a global alignment in the system. This is represented by a color in the majority for higher values of $v_0=0.1$ and $0.5$ as shown in Figs.\ref{fig3}(c-d). However, all through this, the lower magnitude in chirality results in very little orientational difference between the nature of particles and proves insufficient to segregate them, ultimately giving rise to a mixed phase. The mixing can be seen in the mixed nature of both the particle types of colors in Figs.\ref{fig2}(I)(c-d) and minor density differences $\Delta\rho$ in Figs.\ref{fig2}(II)(c-d).\\
As we subject the system to a slightly higher chirality, we observe varying behaviors based on the magnitude of self-propulsion. As the chirality is higher, it dominates over $v_0$ for lower self-propulsion values. The particles are then devoid of any chances to interact with each other, mostly rotating in small circles, and hence remain mixed, reducing the global ordering in the process. This can be seen in Figs.\ref{fig2}(I-II)(e) and Fig.\ref{fig3}(e). On increasing $v_0$, we see that the particles start forming small rotating clusters (as seen in Fig.\ref{fig2}(I)(f)), and each cluster has its local ordering. This is because the chirality is strong enough to create small pockets of rotating clusters, and the self-propulsion causes the particles to separate. On further increase in self-propulsion, the particles traverse larger steps before eventually deflecting from their trajectory, the curvature is strong enough to separate the particles in a short period.\\
\begin{figure*}[!hbtp]
\centering
\includegraphics[width=18cm, height=5cm]{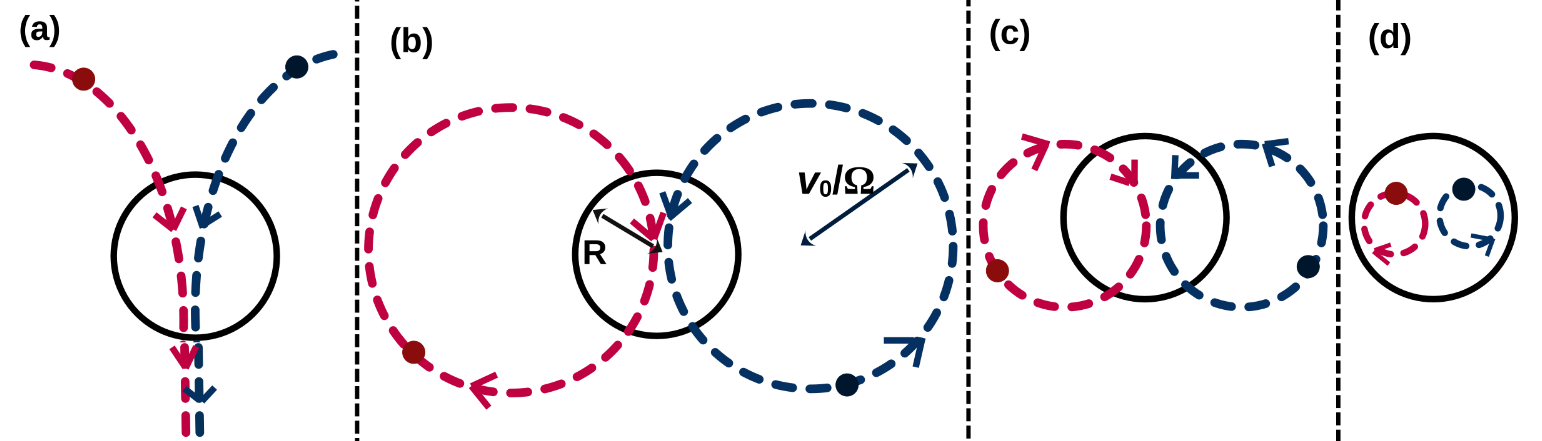}
\caption{The figures (a), (b), (c), and (d) depict the cartoon of the trajectories of one $CW$ (red) and one $ACW$ (blue) chiral active particles. The solid black circle represents the interaction range with a radius  $R=1$, within which the chiral particles interact. The dashed lines show the cartoon of the trajectories of two types of particles, with arrows representing the directions of the velocity of the particles. The dots on the dashed line are the locations of two types of particles at some arbitrary time.
(a)  $v_0=0.5$ and $\Omega=0.01$, the radius of the chiral particle trajectory ($r_0 = \frac{v_0}{\Omega} = 50$) is significantly larger than the interaction range.
 (b) $v_0=0.5$ and $\Omega=0.1$, the trajectories are comparable in size to the interaction radius, resulting in the radius of the trajectory ($r_0 = \frac{v_0}{\Omega} = 5$). (c) $v_0=0.5$ and $\Omega=0.5$, the radius of trajectory $r_0 = \frac{v_0}{\Omega} = 1$ is nearly equal to the   interaction radius. (d) $v_0=0.2$ and $\Omega=0.5$, the particle trajectories have a smaller radius $r_0 = \frac{v_0}{\Omega} = 0.4$ than the interaction radius. }
\label{fig4}
\end{figure*}
From the above discussion, we can conclude that at intermediate values of $(\Omega)$, when the two types of particles undergo phase segregation, the resulting clusters behave more like condensation than traditional phase separation. Notably, as the system size $(L)$ increases, the size of the clusters $l_{sat}$ remains roughly constant, Figs.\ref{fig5}(b,c), while their density rises according to a power-law scaling Fig.\ref{fig5}(a). Specifically, the density difference in condensed  clusters is as $(\Delta \rho^* \sim L^2)$ where $\Delta \rho^*= \{\rho^+({\bf r})-\rho^-({\bf r})\}_{max}$, where $\{...\}_{max}$ means the maximum of local density difference  $\rho^+$ and $\rho^-$. Further, the $\Delta \rho^*$ is obtained by averaging over time in the steady state and over $500$ independent realizations. The variation of $\Delta \rho^* $ as $L^2$ means that the number of particles in each cluster grows much faster than linear dependence, leading to denser structures without an increase in cluster size. In contrast, conventional phase segregation typically involves coarsening, where the cluster size increases with the system size. In our case, the saturation length of the clusters does not change significantly, indicating that the clusters maintain a finite size regardless of the growing system size. Instead, additional particles condense into existing clusters, rapidly increasing density. This behavior aligns with condensate clusters, where a portion of the system's mass condenses into high-density regions, as seen in other active matter systems \cite{wang2024condensation}. Thus, these phase-separated structures are better classified as condensation-driven clusters.\\
Now, as we go to even higher chirality $(0.3)$, smaller regimes of self-propulsion are expected to give rise to mixed phases as rotational tendencies of the particles dominate over their self-propulsion. On increasing self-propulsion, we see that the particles group to form small clusters with local cluster alignment, similar to the case discussed previously. However, there is a subtle difference from the previously discussed case of chirality with high self-propulsion in continuing to increase self-propulsion. Now, the cumulative effects of comparatively higher chirality and self-propulsion $v_0$ enhance the translational motion of the particles and deflect their trajectories, weakening the orientation interaction, in the process, resulting in a reduced density within the cluster (as shown in Fig.\ref{fig2}II(l)). However, chirality keeps the particles in a phase-segregated state and maintains a common orientation with their neighbors, as shown in Fig.\ref{fig3}(l).
As we continue to increase chirality to a value of $0.5$, particles exhibit mixed phases (as seen in Figs.\ref{fig2}(I-II)(m) and Fig.\ref{fig3}(m)) similar to the previously discussed cases for low self-propulsion velocities. Here on increasing self-propulsion, although the particles form local clusters, their enhanced intrinsic rotational dynamics destroy the overall ordering within the cluster, in turn weakening their phase transition and ultimately leading to dilute and expanded clusters (as seen in Figs.\ref{fig2}(I-II)(n-p) and Figs.\ref{fig3}(n-p)).

Based on the range of chirality $\Omega$ and activity $v_0$, we characterize the four cases in the system. We can define the ratio $r_0 = \frac{v_0}{\Omega}$ as the radius of the trajectory of a single particle at the mean-field level. Hence, the ratio determines the curvature of the particle trajectory. The large ratio $r_0$ means almost straight-line motion, and the small ratio $r_0$ means a particle rotating about its axis. The ratio $r_0$ can be compared to the one relevant length scale in the system: the interaction range $R$. Based on this comparison, we can define the four scenarios in the parameter space. \\
{\em Case I: $r_0 >> R$}: In this case, particle trajectory is nearly a straight line within the range of its interaction radius, and two types of particles may not experience their chirality, and for large $v_0$, they will remain mixed and try to move in the same direction (flocking), and a global mixed ordered state will develop in the system. This is the situation for $(v_0, \Omega) = (0.1-0.5,  0.01)$ as can be seen in Figs.\ref{fig2}(I-II)(c-d) and \ref{fig3}(c-d). The cartoon of the trajectory of two types of particles is shown in Fig.\ref{fig4}(a). 

{\em Case II: $r_0 \ge R$}: In this case, although particles maintain nearly the same orientation when lying within the interaction radius, very soon their trajectory gets deflected, they start moving in opposite directions, and they leave the periphery of their previous neighbors and meet with other sets of neighbors. If the new neighbors have the same chirality, the particles will continue moving in the same direction. If not, their paths will be deflected again, leading them to a new location. This search will end with a state where most particles in the neighborhood are of similar chirality. Hence, particles will segregate phases based on their chirality, and a dense condensed phase develops in the system. Each phase segregates cluster maintains ordering within the cluster as can be seen by Figs.\ref{fig2}(I-II)(g-h, l) and \ref{fig3}(g-h, l). The cartoon of the two particles' trajectories and interactions is shown in Fig.\ref{fig4}(b).

{\em Case III $r_0 \lesssim R$}: In this case, the particle's trajectory has a larger curvature, and as soon as they leave the interaction radius, they segregate from each other. However, since the radius of the trajectory is of the order of the interaction radius, small pockets of clusters develop, and condensed clustering is obtained in the system. The mechanism of condensed clustering is the same as explained in {\em Case II} with a key difference that now particles can only go to a distance smaller than that in {\em Case II}; this leads to the formation of small clusters, as can be seen by Figs.\ref{fig2}(I-II)(f,j) and \ref{fig3}(f,j). The cartoon of the trajectory of the two particles and their interaction is shown in Fig.\ref{fig4}(c).

{\em Case IV: $r_0 << R$}: In this situation, particles do not get a chance to align, and chiral nature dominates; they keep rotating about their axis and do not leave their interaction radius and hence remain mixed. As can be seen by Fig.\ref{fig2}(I-II)(i, m-n) and \ref{fig3}(i, m-n). The cartoon of the same is shown in Fig.\ref{fig4}(d). 

\begin{figure}[H]
\centering
\includegraphics[width=8.5cm, height=5.0cm]{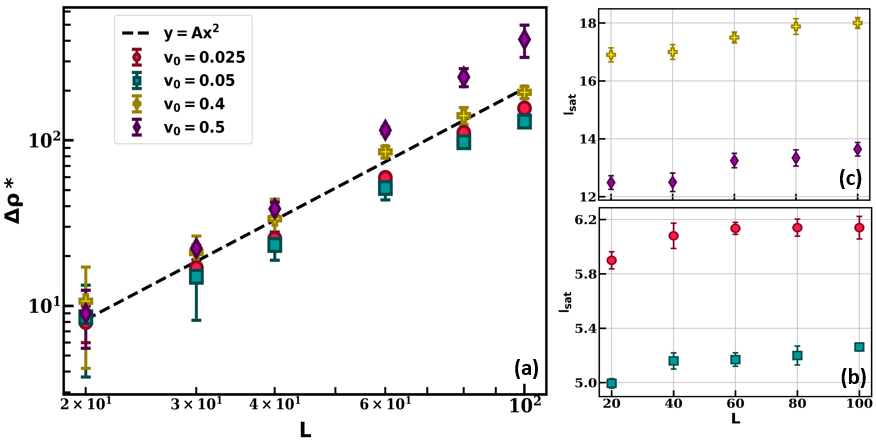}
\caption{(color online)$(a)$ Scaling of condensed polar packet density difference $(\Delta \rho^*)$ with system size $(L)$ for four different parameters in the condensate phase. $(b-c)$ show the saturation length $l_{sat}$ for the same micro and macro condensates respectively.}
\label{fig5}
\end{figure}

\begin{figure*}[!hbtp]
\centering
\includegraphics[width=18cm, height=4cm]{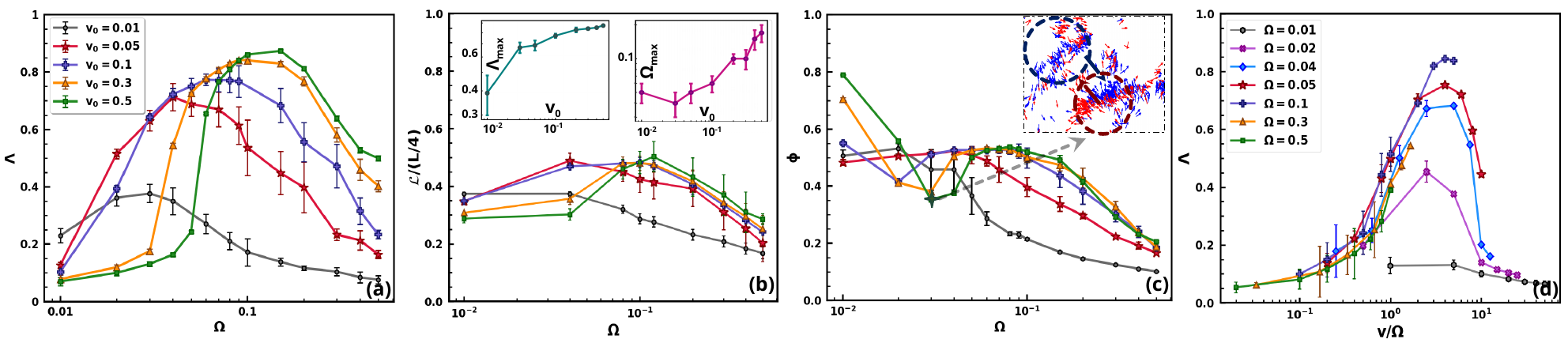}
\caption{(color online) (a) The $\Lambda$  $vs.$ $\Omega$ for different $v_0$. (b) Correlation Length $\mathcal{L}$ $vs.$ $\Omega$ for different $v_0$. Insets (I) $\Lambda_{max}$ $vs.$ $v_0$ and $\Omega_{max}$ $vs.$ $v_0$. (c) The $\Phi$ $vs.$ $\Omega$ for different $v_0$. The snapshot in the corner illustrates the mechanism of mixing, which leads to a dip in $\Phi$. The snapshot is for $v_0 = 0.5$ and $\Omega=0.05$ marked by {\em plus} ${\bf +}$ in the figure. Two highlighted dotted circles (blue/red) show the dominance of negative and positive chiral particles. The small arrows show the direction of orientation of particles, and the thick arrows show the local mean orientation of particles in each circle. The other parameters are the same as in Fig.\ref{fig2}(I). (d) The $\Lambda$ $vs.$ $v_0/\Omega$ for different $\Omega$.}
\label{fig6}
\end{figure*}
We have discussed the system's characteristics based on the snapshots; now, we quantify the system behavior in the steady state by measuring the two order parameters, representing the phase segregation and global orientation order. To explain the observed patterns, as discussed before, we use the Phase segregation order parameter $(PSOP)$ ($\Lambda$) to quantify the degree of mixing and demixing in the system, and the Global order parameter $(GOP)$ ($\Phi$) to quantify the overall order and disorder within the system.\\
The PSOP is defined as:
\begin{equation*} 
\Lambda(t)={\dfrac{|\text{n}(\Omega^+, t)-\text{n}(\Omega^-, t)|}{\text{n}(\Omega^+, t)+\text{n}(\Omega^-, t)}} 
\end{equation*}
where $\text{n}(\Omega^+, t)$ and $\text{n}(\Omega^-, t)$ represent the number of particles with positive and negative chirality, respectively, at time $t$, within the distance $L/4$ (to differentiate the condensed phase segregation in the system. We also checked for other distances, $L/16$, $L/8$, etc., the behavior of $\Lambda$ remains the same.)  Further, we take the mean of $\Lambda(t)$ over the steady state when the $\Lambda(t)$ remains statistically similar over time. The mean is calculated over time in the steady state and over $10$ independent realizations. \\
To further quantify the clustering and phase segregation among the particles, we can also use the local density difference of the two types of particles as defined before and shown in Fig.\ref{fig2}(II). Here, we use the local density difference $\Delta \rho({\bf r}, t)$ at different times in the steady state. Using the local density difference $\Delta \rho({\bf r}, t)$, we calculated the two-point normalized correlation function $C(r, t) = <\Delta \rho({\bf r}_0, t) \Delta \rho({\bf r} + {\bf r}_0, t)>$, where the mean is over all the points ${\bf r}_0$, directions, different times in the steady state, and over $10$  ensembles. The correlation function $C(r, t)$ will depend on the two parameters $v_0$ and $\Omega$ in the steady state. We further define the correlation length $\mathcal{L}(v_0, \Omega, t)$ from the distance the correlation function reaches $10\%$ of its value from the first data point.  \\
In  Figs.\ref{fig6}(a-b) we show the plot of  $PSOP$ and $\frac{\mathcal{L}(v_0, \Omega)}{L/4}$ $vs.$ $\Omega$ for different $v_0 \in (0.01, 0.5)$ respectively. Very clearly, both show the non-monotonic trend as a function of chirality. 
For small activity $v_0 = 0.01$, particles have very weak self-propulsion speed, and for any finite chirality, the rotational nature dominates, and both $PSOP$ and $\mathcal{L}(v_0, \Omega)$ remain low with a small peak at $\Omega \simeq 0.02$. For very high $\Omega$, particles mainly remain stuck to their location and perform a circular motion. Hence, they do not get the chance to move out and phase segregate. For moderate $\Omega$'s close to $0.02$, chirality and activity are in comparison, and they have some tendency to form small clusters. On increasing activity, initially, the $PSOP$ and length $\mathcal{L}$ increase, and then again, for larger chirality, it shows the monotonic decay to small values. The $PSOP$ and $\mathcal{L}$ show the gradual increment on the increasing $v_0$; it reaches a maximum at $\Omega_{max}(v_0)$. The $\Omega_{max}$ shifts towards the larger $\Omega$ on increasing $v_0$, although the shift is weak, as shown in Fig.\ref{fig6}(b) (inset II). 
For activity ranging from medium to high ($v_0 > 0.05$), increasing the angular velocity $\Omega$ keeps the system mixed for very small $\Omega$, and both the $\Lambda$ and $\mathcal{L}$ are low, indicating a lack of phase segregation. As the angular velocity $\Omega$ continues to increase, there is a bias in particle orientation, leading to an increase in phase segregation. Consequently, the $\Lambda$ and $\mathcal{L}$ increase as the system and reaches a maximum at $\Omega_{max}$ and maximum phase segregation occurs with particles of different chiralities segregating into dense clusters, as depicted in Figs.\ref{fig2}(I-II)(g-h, l). However, as $\Omega$ continues to increase, the particles get segregated beyond $\Omega_{max}$. Still, their trajectory is small and segregated in a sparser form, causing the $\Lambda$ and $\mathcal{L}$ to decrease gradually. Also, increasing activity $v_0$ leads to an increase in the phase segregation, as can be seen by the maximum value of $\Lambda_{max}$ increasing with $v_0$. The plot of $\Lambda_{max}$ $vs.$ $v_0$ is shown in Fig.\ref{fig6}(inset I). For lower values of $v_0$ $v_0 \leq 0.05$, the system does not exhibit a phase segregated state, rather multiple small clusters are formed as can be seen by snapshots Figs.\ref{fig2}(I-II)(f,j), the small length $\Lambda$ and $\mathcal{L}$ values in Figs.\ref{fig5}(a-b). Consequently, the $\Lambda$ ($\mathcal{L}$) does not increase significantly as it does for higher $v_0$ values.\\ 
To further analyze the phase segregation on the variation of self-propulsion speed $v_0$. The effect of self-propulsion speed for different chirality is shown by the scaled plot as shown in Fig.\ref{fig6}(d). We plot $\Lambda$ $vs.$ the scaled variable $v_0/\Omega$ for various fixed values of $\Omega$. For moderate to high chirality, all curves collapse onto a single master curve, revealing that the effective trajectory radius $r_0 = v_0/\Omega$ is the key factor controlling phase segregation. A peak in $\Lambda$ occurs consistently near $v_0/\Omega \sim 1$, where the turning radius becomes comparable to the interaction range $R$, in agreement with the clustering mechanisms described in Cases II and III. For very small $\Omega$ (e.g. $0.01$ and $0.02$), this collapse fails, and $\Lambda$ remains low in all $v_0/\Omega$, indicating that particles behave nearly as straight movers and stay mixed. This confirms that $v_0/\Omega$ effectively captures the onset and strength of chirality-driven segregation in the system.\\
We also examine the Global Order Parameter ($\Phi$) to understand the nature of the orientational ordering in our system. The $\Phi$ is determined by calculating the absolute value of the average normalized velocity of all the particles in the system:
\begin{equation*} 
\Phi(t)= \dfrac{1}{N v_0}\bigg|\displaystyle\sum_{i=1}^{N}v_i(t) \bigg|
\end{equation*}
Further, the $\Phi(t)$ is averaged over the times in the steady state and ensembles to get the mean order parameter $\Phi$. The $\Phi$ is approximately equal to one when all particles move coherently in the same direction, indicating a highly ordered state. Conversely, the $\Phi$ approaches zero when the motion of individual particles is random, reflecting a disordered state.\\
For activities in the medium to high range ($v_0 \geq 0.1$), the global order parameter ($\Phi$) initially decreases from its maximum value, as shown in Fig.\ref{fig6}(c). At low values of $\Omega$, $\Phi$ is at its peak because the particles exhibit a low tendency to rotate and remain mixed and move in the same direction. Demixing creates a deflection in their trajectory and decreases in $\Phi$. The $\Phi$ also increases as $v_0$ increases while $\Omega$ remains low. This can be observed in the orientation snapshots of particles in Fig.\ref{fig3}(c-d), which shows that at $\Omega = 0.01$, as $v_0$ increases, there is a higher likelihood that all particles within a cluster align in the same direction, resulting in a more uniform color within the cluster. This indicates that the particles have a reduced tendency to rotate, leading to more coherent movement in a single direction. Consequently, the system achieves a highly ordered state, which is reflected in Fig.\ref{fig5}(c), where the $\Phi$ reaches its maximum when $v_0$ is at its highest value under low $\Omega$ conditions.\\
As $\Omega$ increases, the bias in particle orientation becomes stronger. This causes particles to rotate in different directions within a mixed state, forming distinct clusters that contain both types of particles. In this mixed state, some clusters are dominated by particles with positive chirality, causing them to rotate clockwise $(CW)$. In contrast, others are dominated by particles with negative chirality, leading them to rotate anticlockwise $(ACW)$ Fig.\ref{fig6}(c). However, these clusters are not perfectly ordered due to the mixed nature of the particles. As particles continuously join or leave the clusters, they do not rotate at a fixed radius, leading to frequent collisions between the mixed clusters. This behavior increases the overall randomness and disorder within the system, resulting in a sudden decrease in the $\Phi$ value, marking a significant change. This behavior is shown by particles when $v_0 \geq 0.07$.  
At lower values of $v_0$, there is no sudden decrease in the $\Phi$; instead, it decreases slowly and becomes approximately constant at large $\Omega$, indicating that the particles are moving very slowly and there is no collision between particle clusters. The mechanism of collision is shown in the zoomed snapshot of the particles' position with the orientation indicated by arrows, Fig.\ref{fig5}(c).\\
However, particles eventually fully segregated phase at higher $v_0 (\geq 0.1)$ and higher $\Omega$. At high velocities, a second maximum of $\Phi$ is observed at $\Omega$ around $0.1$. Beyond this point, as $\Omega$ further increases, $\Phi$ gradually decreases and stabilizes at an approximately constant value for a given $v_0$. In particular, the first maximum of $\Lambda(\mathcal{L})$ and the second maximum of the $\Phi$ occur at the exact value of $\Omega$ for a particular velocity, signifying the appearance of a dense condensed phase with global rotation of particles within their clusters in the system. 

\subsection{Dynamics of the system}
\label{subsec:Dynamic}
We have now discussed the system's characteristics regarding steady-state ordering and density clusters for different activities and chirality. Now, we will explore the dynamics of two types of particles in the system.
\begin{figure}[H]
\centering
\includegraphics[width=8.5cm, height=7.5cm]{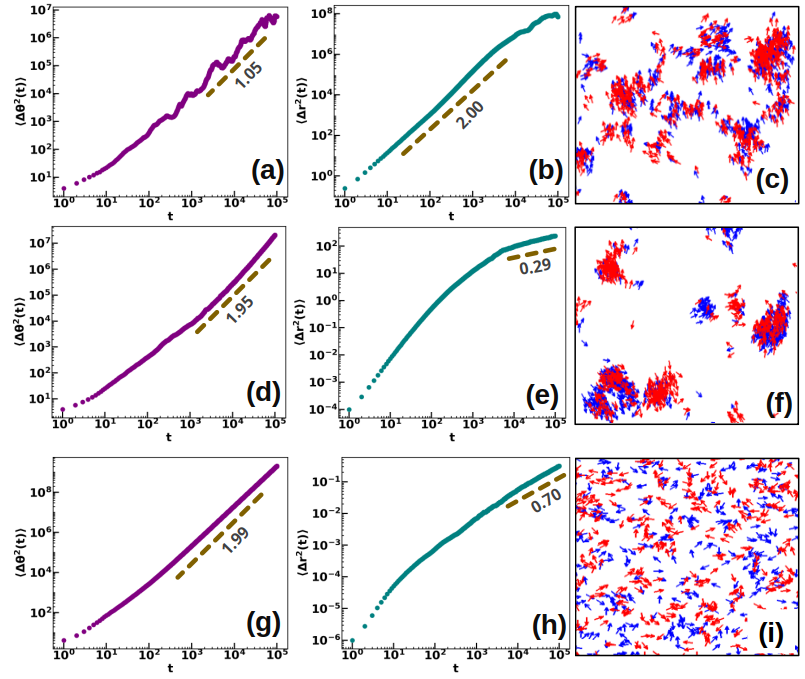}
\caption{$(a,d,g)$ {\em MSAD} $\langle \Delta \theta^2 (t)\rangle$ $vs.$ $t$ and $(b,e,h)$ {\em MSD} $\langle \Delta r^2 (t)\rangle$ $vs.$ $t$ plots for the parameters where we find mixing of two types of particles and $(c,f, i)$ show the snapshots for the same parameters. The arrows in the snapshots  represent the orientation of particles and colors have the same meaning as in Figs.\ref{fig2}I $(a,b,c)$,$(d,e,f),$ and $(g,h,i)$ are for the set of parameters $(v=0.5, \Omega=0.01), (v=0.01, \Omega=0.05)$ and $(v=0.001, \Omega=0.5)$.}
\label{fig7}
\end{figure}
We define the {\em Mean Square displacement $(MSD)$ and Mean square angular displacement $(MSAD)$} as $\langle \Delta r^2 \rangle (t) = \dfrac{1}{N}\langle \sum_{i=1}^{N} |\textbf{r}_i(t+t_0)-\textbf{r}_i(t_0)|^2\rangle$ and $\langle \Delta \theta^2 \rangle (t) = \dfrac{1}{N}\langle \sum_{i=1}^{N} |\theta_i(t+t_0)-\theta_i(t_0)|^2\rangle$ respectively, where $\textbf{r}_i(t)$ and $\theta_i (t)$ are the position and angle of the $i^{th}$ particle at time $t$, $t_0$ represents the reference time of the measurement. Here, $\langle.....\rangle$ indicates averaging over different values of $t_0$ in the steady-state trajectory. Due to periodic boundary conditions in both directions, the motion happens without any confinement, and $MSD$ has no upper bound. Hence, the behavior of $MSD$ and $MSAD$ with respect to time will show the translational and angular dynamics of the particles for different parameters. Generally, $MSD$ and $MSAD$ follow a power-law dependence on $t$. $\langle \Delta r^2 \rangle\sim t^\alpha$ and $ \langle \Delta \theta^2 \rangle \sim t^\beta$, where the exponent $\alpha = 1$ corresponds to diffusive motion, $\alpha<1$ to subdiffusive motion, and $\alpha>1$ to superdiffusive motion, while for ballistic motion, one has $\alpha=2$. Similarly, $\beta=1$ represents the random reorientation, and $\beta=2$ shows the persistent rotation of particles. In Figs.\ref{fig7}(c,f,i), we again show the snapshots of the system for three different combinations of activity and chirality $(v_0, \Omega) = (0.5, 0.01), (0.01, 0.05)$ and $(0.001, 0.5)$ respectively. 
The three cases are chosen as follows: (a) When the trajectory of the particle is a nearly straight line (Figs.\ref{fig2}(I-II)(d) \ref{fig3}(d)). This case shows mixed flocking behavior, as seen by the ballistic translational motion through $MSD$ and random reorientation through $MSAD$, as shown in Figs.\ref{fig7}(a-c). However, at late times, the $MSD$ shows saturation due to finite chirality in the system. This clearly explains the enhanced global order parameter $\Phi$ and small $PSOP$ for this set of parameters. In Fig.\ref{fig7}(f), we show the snapshot for $v_0=0.01$ and $\Omega = 0.05$, where chirality starts to dominate, and particles start moving in a small radius circle shown by $MSAD$ ballistic motion Fig.\ref{fig7}(d). Since the $r_0 < R$, the particles can mix and rotate in small clusters. The $MSD$ shows the trapped motion at late times due to the rotational dynamics, as shown in Fig.\ref{fig7}(e). Now, as we decrease $v_0=0.001$ and increase $\Omega=0.5$ as shown in the snapshot in Fig.\ref{fig7}(i), chirality is dominant, and particles mainly rotate about their axis as shown by the ballistic behavior of $MSAD$ Fig.\ref{fig7}(g) without interacting with their neighbors. Hence, the system remains in a homogeneous phase with a minimal magnitude of $MSD$ compared to Fig.\ref{fig7}(h). The dynamics of the particle are very different for the two cases Figs.\ref{fig7}(e) and (f). In Fig.\ref{fig7}(e), when both activity and chirality have moderate values, but activity dominates, particles can form mixed oriented clusters with a rotational tendency, leading to a very small $MSD$ exponent $\alpha << 1$. On the other hand, for Fig.\ref{fig7}(h), when activity is minimal and chirality is dominating, particles mainly rotate about their axis with small random displacement from their mean position, resulting in a small magnitude of $MSD$ and subdiffusive dynamics in the system. 
\subsection{Phase diagram}
\label{subsec:Phase_diag.}
\begin{figure}[hbt]
\centering
\includegraphics[width=8.5cm, height=7cm]{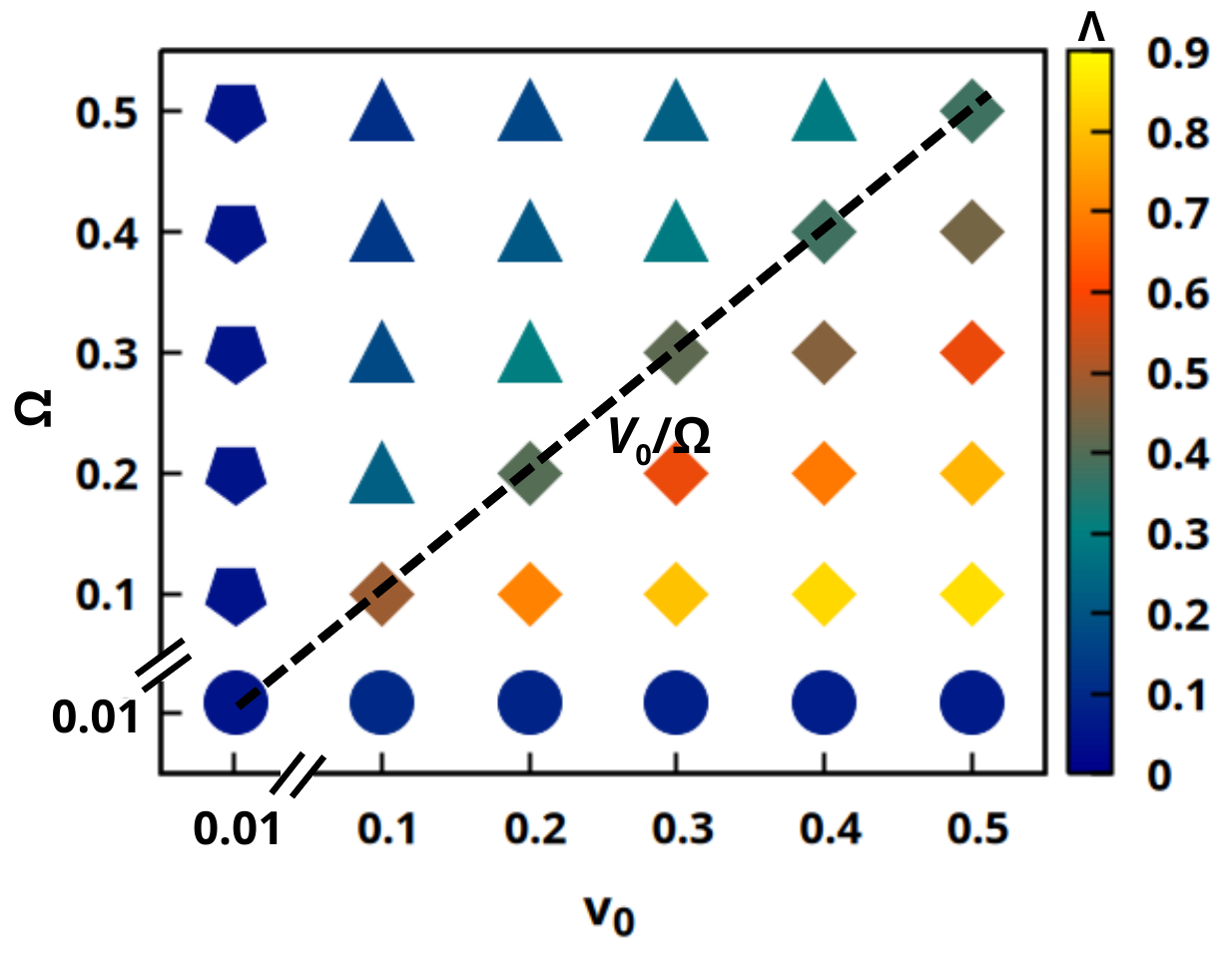}
\caption{(color online)Phase Diagram for different phases in $(\Omega,v_0)$ plane. The circle shows the mixed flocking $(MFP)$, the pentagon shows the homogeneous rotating $(HRP)$, and the diamond shows the condensed phase $(CP)$. The color bar shows the magnitude of the $PSOP$ $\Lambda$. The black dotted line shows the $\frac{v_0}{\Omega} = 1 = R$ line}.
\label{fig8}
\end{figure}
We have analyzed the steady-state features of the system concerning activity and chirality, and we drew the phase diagram in the plane of activity $v_0$ and chirality $\Omega$ as shown in Fig.\ref{fig8}.  The colors in each symbol represent the magnitude of $PSOP$, and the different symbols represent three phases of the system based on relative activity and chirality. We have already discussed the detailed characteristics of the system for these different phases. Now, based on their characteristics, we name and summarize them here. For very small chirality ($\Omega \le 0.01$) and across the entire range of activity, the particles exist in a mixed state with small correlation length and $PSOP$ and flock in a common direction with moderate to large values of $GOP$. We refer to this phase as the mixed flocking phase $(MFP)$ as shown by the circles in Fig.\ref{fig8}. When the activity is small ($v_0 < 0.001$) and there is finite chirality, the particles rotate about their axis, and both types of particles remain mixed, resulting in small values of $GOP$, $PSOP$, and $\mathcal{L}$. The pentagons show such a phase in Fig.\ref{fig8}, and we named this phase and homogeneous rotating phase $(HRP)$.
 We have observed interesting phases emerging when activity and chirality compete, where particles undergo segregation and exhibit condensation rather than conventional phase separation. Within this condensed phase $(CP)$, the nature of condensation depends on the ratio $v_0/\Omega$. When $r_0 >1$, particles maintain alignment over longer distances before their trajectories curve significantly, leading to dense condensation, where clusters absorb a substantial fraction of particles and exhibit strong localization. In contrast, when $r_0<1$, the strong rotational bias restricts long-range alignment, resulting in weak condensation, where clusters form but remain more loosely structured and dynamic. As shown in Fig.\ref{fig8} $PSOP$.\\
Till now, we have discussed the steady properties of different phases. 
\begin{figure*}[!hbtp]
\centering
\includegraphics[width=16cm, height=10.5cm]{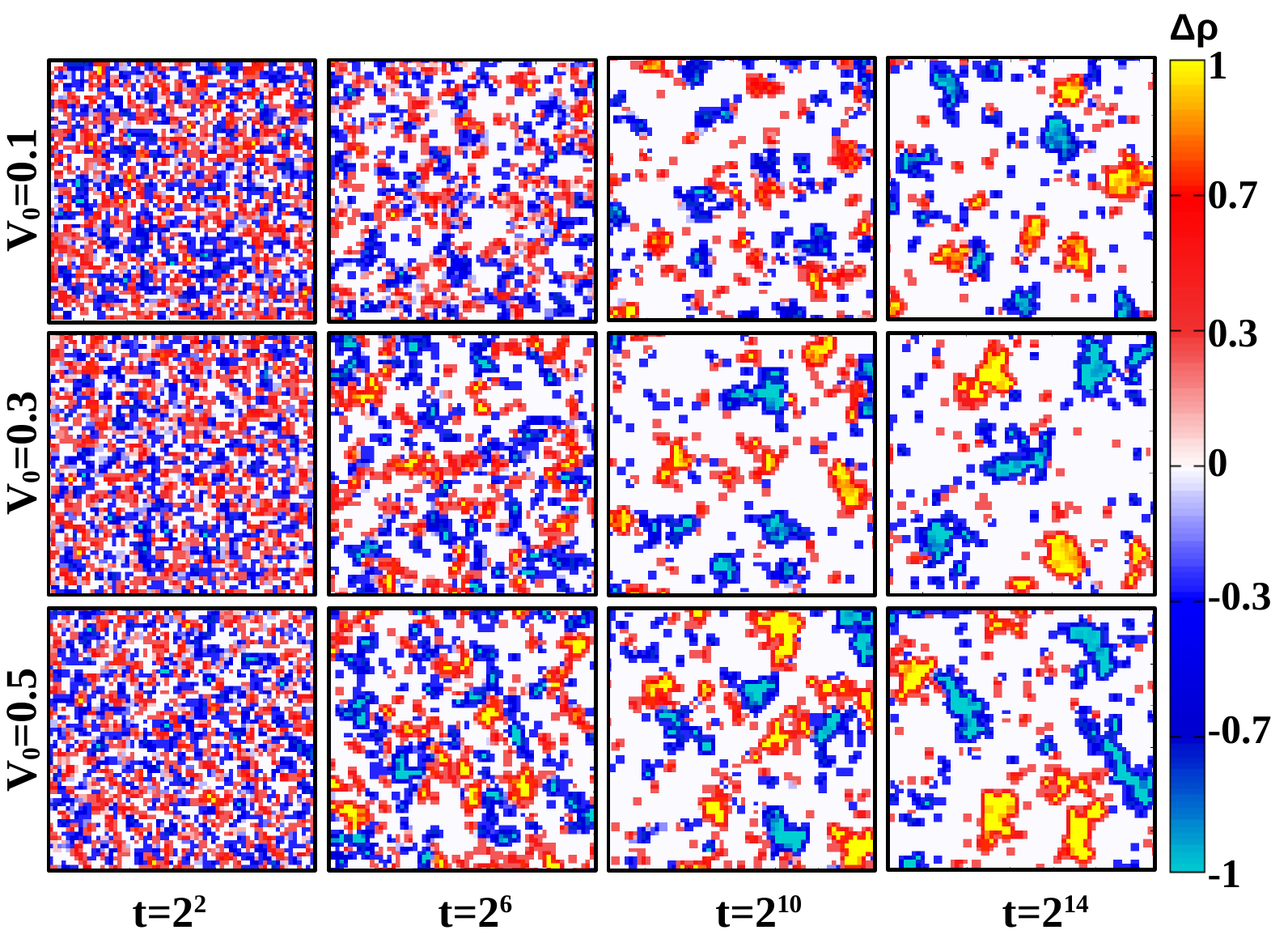}
\caption{(color online) Shows the time-dependent density difference $\Delta \rho$ snapshots for different $(v_0=0.1,0.3,0.5)$ and $\Omega =1$  for $t=2^2, 2^6, 2^{10}$ and $2^{14}$ and color palette is same as in Fig.\ref{fig2}(II).}
\label{fig9}
\end{figure*}
\begin{figure}[hbt]
\centering
\includegraphics[width=8.5cm, height=3.5cm]{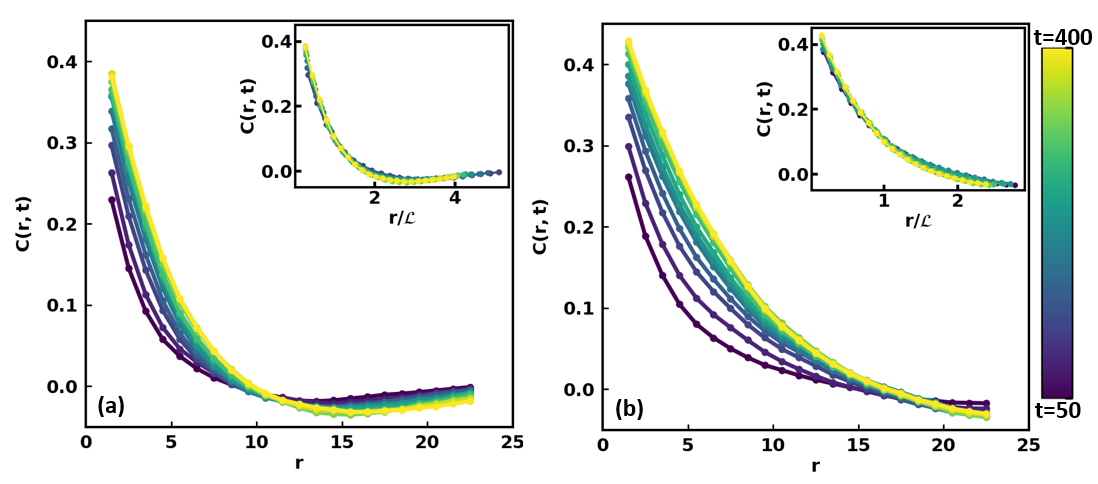}
\caption{(color online) $(a)$ $v_0 = 0.3$ and $(b)$ $v_0 = 0.6$, is two-point correlation function $C(r,t)$ $vs.$ $r$ at different times for chirality $\Omega = 0.1$. {\em Insets}: the scaled $C(r,t)$ $vs.$ scaled distance $r/\mathcal{L}$  for all five cases.}
\label{fig10}
\end{figure}
\begin{figure}[hbt]
\centering
\includegraphics[width=8.5
cm, height=4cm]{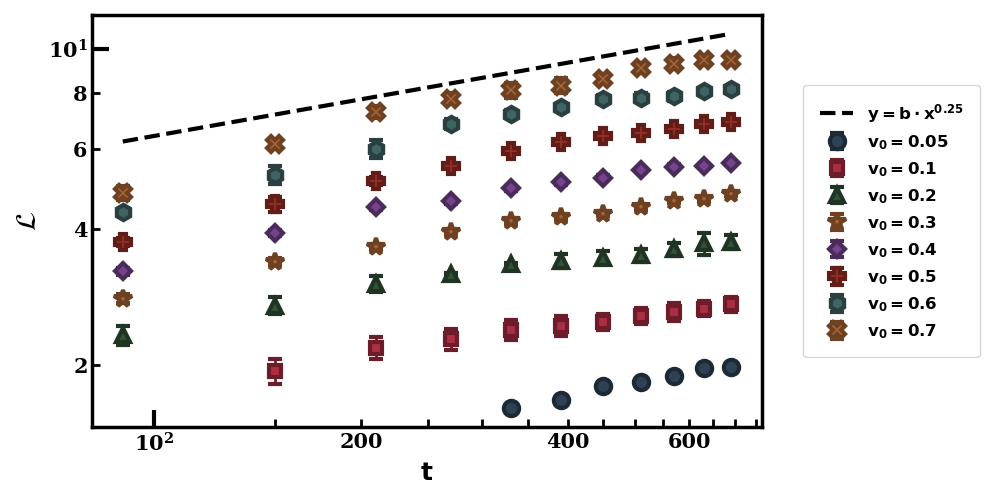}
\caption{(color online) Plot of correlation length  $\mathcal{L}$ $vs.$  $t$ in the growing stage on  $\log$-$\log$ scale for different $v_0$ at $\Omega=0.1$ values.}
\label{fig11}
\end{figure}
\begin{figure*}[!hbtp]
\centering
\includegraphics[width=18cm, height=4cm]{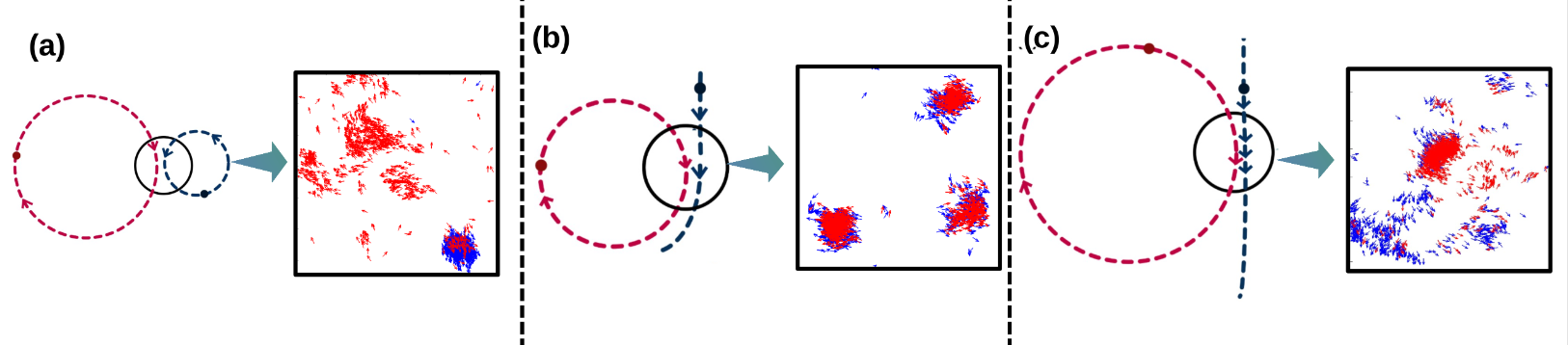}
\caption{(a) Cartoon of the trajectories of two particles for the asymmetric combinations of chirality. The color code and keys are the same as for Fig.\ref{fig4}. The snapshots on the right side of each panel show a part of the system for the same set of parameters. The keys of the snapshots are the same as for Figs.\ref{fig6}.$(a$-$c)$ is for the set of parameters $(v_0,\Omega^+,\Omega^-) = (0.5, 0.1, -0.3), (0.5, 0.1, 0)$ and $(0.3, 0.1, 0)$ respectively.}
\label{fig12}
 \end{figure*}
\subsection{Kinetics of condensing clusters}
\label{subsec:Kinetics}
To understand the kinetics, we first observe the system's time evolution. In Fig.\ref{fig9} we show the snapshot of the local density difference $\Delta \rho({\bf r})$  at four different times $t=2^2$, $2^6$, $2^{10}$ and $2^{14}$ for $(v_0)= ((0.1),(0.3),(0.5))$ and $\Omega=1$. The choice of parameters is only the combinations of activity and chirality where we get $(CP)$. With time, the phase segregation among the two types of particles increases, as can be seen by the appearance of bright spots in the last two columns of Fig.\ref{fig9}. Also, the size of colored regions increases with time. \\
Furthermore, we show the plot of the two-point correlation function of the local density difference as defined previously in Subsection \ref{subsec:characteristion}. In Figs.\ref{fig10}(a,b), we show the plot of $C(r,t)$ for two different $(v_0=0.3,0.6)$. With time, the correlation increases, and we calculate the correlation length $\mathcal{L}(v_0, \Omega, t)$. In the insets of Figs.\ref{fig10}(a,b), we show the scaled correlation function with scaled distance $r/\mathcal{L}$ for different twelve times between $t=50$ to $400$. The collapse of data for the scaled plots will represent the dynamic scaling in the system. %We find the scaling improves as we go for higher $v_0$. 
Fig.\ref{fig11} shows the variation in the length of the correlation $\mathcal{L}(v_0, \Omega, t)$ {\em versus} time $t$ for different combinations of $(v_0=0.05,0.1,0.2,0.3,0.4,0.5,0.6$ and $0.7)$ at $\Omega=0.1$. For all cases, initially, $\mathcal{L}$ grows with time with a power $t^{\gamma}$, with $\gamma \sim 0.25$, and at late times the growth of $\mathcal{L}(v_0, \Omega, t)$ slows as the system approaches a steady state.   
\subsection{Asymmetric mixture}
\label{subsec: Asymmetric}
Till now, our focus has been on the symmetric mixture of two types of chiral particles. Now, we look for a case where the magnitude of the chirality of two particle types differs. Hence, we call it an asymmetric mixture. In this case,  again $50\%$ of the particles exhibit clockwise $(CW)$ chirality, while the remaining $50\%$ exhibit anticlockwise $(ACW)$ chirality, but the particles' chirality magnitudes are not equal, indicating that the orientation of the two types of particles is not exactly opposite to each other. We will discuss the three combinations of chirality and activity. (i) As shown in Fig.\ref{fig12}, $ACW$-chiral particles ($\Omega^{-}=0.3$) exhibit higher chirality compared to $CW$-chiral particles ($\Omega^{+}=0.1$). In this case, chiral particle trajectories are a mixture of {\em Case-II} and {Case-III} as shown in Figs.\ref{fig4}(b, c), and previously discussed in subsection \ref{subsec:characteristion}. Due to their smaller curvature, blue $(ACW)$ particles spend more time in the interaction range than red $(CW)$  particles with larger curvature.\\
\begin{figure}[hbt]
\centering
\includegraphics[width=6
cm, height=5cm]{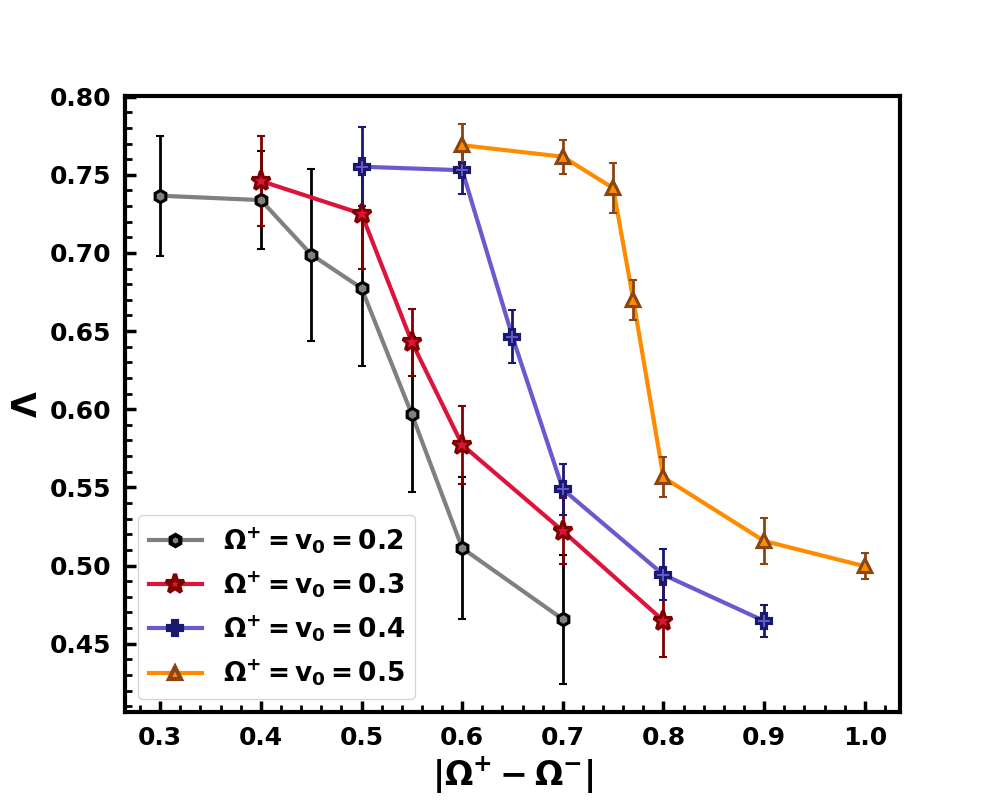}
\caption{(color online) $\Lambda$ $vs.$ $|\Omega^{+} - \Omega^{-}|$ for different values of $v_0 = \Omega^{+}$ (where, $|\Omega^{+} - \Omega^{-}|=|\Omega^{+}|+|\Omega^{-}|$). Each curve corresponds to a system where both the self-propulsion speed and the chirality of one species are fixed and equal. In contrast, the chirality of the second species $\Omega^{-}$ is varied to introduce asymmetry.}
\label{fig13}
\end{figure}
 Consequently,  particles with negative chirality form one strong cluster and rotate due to their chirality. In contrast, positive chiral particles rotate in a dispersed manner in the system, as shown in the snapshot in Fig.\ref{fig12}(a).
The other two cases are (ii) when the one-type particle is achiral $\Omega^{-} = 0$ and $\Omega^{+} = 0.1$ and $v_0 = 0.3$ for both types of particles. In Fig.\ref{fig12}(b), we show the trajectory of two particles and the snapshot of the system at some fixed time in steady state.
  In this scenario, when activity is  $v_0=0.3$, both particles form the mixed cluster, and the mixture shows chiral behavior. This happens because if the hopping step of particles is smaller than the $R$, then particles have a higher tendency to remain within the interaction range for some time and get a chance to interact with the mixture. Also, the curvature of the chiral particles is larger, resulting in them following the circular trajectory of the chiral particles, and the non-chiral particles (blue) rotate due to the chirality of the positive chiral particles, as shown in the snapshot of Fig.\ref{fig12}(b). However, for the third case (c), when $v_0 = 0.5$, the increased step size of particles causes non-chiral particles to quickly leave the vicinity of chiral particles, as shown in Fig.\ref{fig12}(c). To quantify the phase segregation among two types of particles for the asymmetric case, in Fig.\ref{fig13} we show the plot of phase separation order parameter $\Lambda$ $vs$ difference in the chirality of two types of particles $|\Omega^{+}-\Omega^{-}|$. The $|\Omega^+|$ is kept fixed to $v_0$ and $v_0$ is varied between $(0.2-0.5)$. Hence, for one type of particle trajectory, the radius is roughly the same as the interaction radius. The chirality of the second type particles $\Omega^-$ varies from small to large values. When $|\Omega^{-}| < |\Omega^{+}|$, then both types of particles phase segregate and the system configuration is similar to that shown in Fig.\ref{fig12}(a). The two types of particles phase separate into dense and dilute clusters. As we increase $|\Omega^-|$, the second type of particles starts to move in smaller circles, and the $\Lambda$ shows a dip. At larger $|\Omega^-|$, the second type of particles almost show confined motion to their location, and the system shows the partial phase segregation with $\Lambda ~ 0.5$ value. The activity $v_0$ also plays a crucial role in the phase segregation. The moderate activity values like $v_0 = 0.2$ and $0.3$ revealed that increasing chirality asymmetry leads to a steady decrease in $\Lambda$. This occurs due to differences in rotational dynamics between particle types, which disrupt their alignment. At $v_0 = 0.2$, the system transitions through Cases II and III, where particles initially segregate due to favorable interactions. However, as chirality mismatch increases, the system shifts to Case IV, where particles primarily rotate in place, diminishing their ability to align and significantly reducing $\Lambda$. As activity increases further, particularly at $v_0 = 0.4$ and $0.5$, most of the system remains in the Cases II and III regimes. Enhanced velocities lead to larger effective trajectories, allowing for greater movement before deflections occur. This increased movement boosts the chances of particles aligning with similar neighbors, promoting phase segregation. Consequently, even with asymmetric chirality, the $\Lambda$ remains significant at higher velocities.
\section{Discussion}\label{secIV}
We comprehensively studied a mixture of two types of chiral active particles. Most of our focus was on the symmetric mixture, where the two types of particles have the same magnitude of chirality but different signs. One type of particle rotates clockwise $(CW)$, and the other rotates anticlockwise $(ACW)$. We used activity $v_0$ and magnitude of chirality $\Omega$ as two control parameters in our model. The density of both types of particles was kept the same in the system. The magnitude $\Omega$ measures the rotational nature of particles and the difference between the two types of particles. For small $\Omega$, the difference in chirality between the two types of particles is slight, causing them to remain in a mixed phase and move together in the same direction. Hence, the system is a mixed flocking phase $(MFP)$. For small activity and finite chirality, chirality dominates, and the particles remain mixed, rotating about their axes without much translational motion. This phase is named the homogeneous rotating phase $(HRP)$. %\textcolor{blue}
Interestingly, the system exhibits a condensed phase when activity and chirality are comparable. When $r_0>1$, particles maintain alignment over longer distances before their trajectories curve significantly, leading to dense condensation, where clusters absorb a substantial fraction of particles and exhibit strong localization. In contrast, when $r_0<1$, the strong rotational bias restricts long-range alignment, resulting in weak condensation, where clusters form but remain more loosely structured and dynamic. 

We also studied the kinetics of phase segregation for the condensation clusters. We observed that the correlation length grows algebraically with time,  with a growth exponent similar to the usual growth law for active systems \cite{pattanayak2021domain,redner2013structure}, and the system exhibits good dynamic scaling. Although the system shows the three phases concerning activity and chirality and interesting dynamic patterns, we discussed the explanation of the three phases in detail with the help of a simple mean-field understanding using the radius of the particle trajectory and the interaction radius.\\
Recent studies~\cite{ai2023spontaneous, ai2018mixing, li2024spontaneous} have shown that phase segregation can arise in binary mixtures of chiral active particles with explicit excluded-volume interactions. In such systems, steric repulsion plays a crucial role in driving motility-induced phase separation, even in the absence of chirality or alignment. In contrast, our model considers point-like particles without volume exclusion, where dynamic clustering and phase segregation emerge purely from the interplay of velocity alignment and opposite chirality. This allows us to isolate chirality-induced effects and characterize the resulting condensed phases as a consequence of curvature-driven self-organization, rather than steric self-trapping. Furthermore, we explored asymmetric mixtures for a few combinations of activity and chirality, in contrast to the symmetric mixtures, where the system exhibits condensed phase segregation and rotating clusters. We considered two cases where the chirality of one type of particle is smaller than the other, and two different activities are chosen, $v_0 = 0.3$ and $0.5$. We found two phases: mixed rotating and rotating flocking phases.\\
Our study provides a deep understanding of the different phases of the mixture of chiral particles. The various phases we observed here can help us understand the patterns and dynamics observed in chiral active particles. The present study is limited to the point particles with a uniform magnitude of chirality. 
It would be interesting to study the system with inhomogeneous chirality.\\ 
\begin{center}
\textbf{Author Contribution}
\end{center}
The first author, DK, conducted the simulations, wrote the code, performed the result analysis, and generated the data. DK also wrote the initial draft of the manuscript. SM designed the model, assisted with coding, contributed to the discussion of the results, and prepared the final version of the manuscript.
\begin{center}
\textbf{Conflict of interest}
\end{center}
There are no conflicts of interest to declare.
\begin{center}
\textbf{Acknowledgment}
\end{center}
The support and the resources provided by PARAM Shivay Facility under the National Supercomputing Mission, Government of India, at the Indian Institute of Technology, Varanasi, are gratefully acknowledged by all authors. SM thanks SS Manna
for useful discussions. SM thanks DST-SERB India, CRG$/2021/006945$, and MTR$/2021/000438$ for financial support; DK acknowledges UGC India for financial support. DK and SM thank the Centre for Computing and Information Services at IIT(BHU), Varanasi.            
 \bibliography{citation}
            
\end{document}